\documentclass[aps,prl,twocolumn,floatfix,10pt]{revtex4-2}

\usepackage{amsmath}
\usepackage{amssymb}
\usepackage{graphicx}
\usepackage{hyperref}
 \usepackage{booktabs}
\usepackage{multirow}
\usepackage{lipsum}
\usepackage{natbib}
\usepackage[english]{babel}

\hypersetup{
    colorlinks=true,
    linkcolor=blue,
    citecolor=blue,
    urlcolor=blue
}

\begin{document}

\title{Hilbert space fragmentation in driven-dephasing Rydberg atom array}

\author{Tianyi Yan}
\author{Chun Hei Leung}
\author{Weibin Li}
\address{School of Physics and Astronomy and Centre for the Mathematics and Theoretical Physics of Quantum Non-equilibrium Systems, University of Nottingham, Nottingham NG7 2RD, United Kingdom}

\date{\today}

\begin{abstract}
We investigate the onset and mechanism of Hilbert space fragmentation (HSF) in a chain of strongly interacting Rydberg atoms subject to local dephasing. It is found that the emergence of multiple long-lived metastable states is fundamentally tied to HSF of the driven-dephasing Rydberg atom system. We demonstrate that the manifesting HSF is captured by a dephasing PXP model that supports multiple degenerate zero modes. These modes form disconnected, block-diagonal subspaces of maximally mixed states, which consist of many-body spin states sharing the same symmetry. A key result is the identification of the underlying symmetry in the HSF, where conserved quantities in each subspace are defined by the consecutive double excitation addressing operator. Moreover, we show explicitly that the number of the fragmented Hilbert space grows exponentially with the chain length, following a modified Fibonacci sequence. Our work provides insights into many-body dynamics under dynamical constraints and opens avenues for controlling and manipulating HSF in Rydberg atom systems.  
\end{abstract}

\maketitle
\textit{Introduction}--
Dynamical constraints can break Hilbert spaces of a many-body system into disconnected subsectors~\cite{sala_ergodicity_2020,khemani_localization_2020,royStrongErgodicityBreaking2020}, leading to Hilbert space fragmentation (HSF)~\cite{yangHilbertSpaceFragmentationStrict2020,hahnInformationDynamicsModel2021,moudgalyaHilbertSpaceFragmentation2022,moudgalyaQuantumManybodyScars2022}. Dynamics confined in a subspace depends sensitively on the symmetry and initial states. A celebrated paradigm is the PXP and related models~\cite{PhysRevB.63.224401,PhysRevB.69.075106,PhysRevLett.106.025301,khemani_signatures_2019,choi_emergent_2019,lin_slow_2020,verresen_prediction_2021,karle_area-law_2021,surace_exact_2021,su_observation_2023,ivanov_volume-entangled_2025,kerschbaumer_quantum_2025}, which has been demonstrated with Rydberg atoms~\cite{bernien_probing_2017,bluvstein_controlling_2021}. In the Rydberg atom system, the interaction induced dipole blockade generates the dynamical constraint~\cite{saffmanQuantumInformationRydberg2010,shaoRydbergSuperatomsArtificial2024a}, which breaks ergodicity~\cite{palmerModelsHierarchicallyConstrained1984,fredricksonKineticIsingModel1984} and eigenstate thermalization hypothesis~\cite{PhysRevA.43.2046, PhysRevE.50.888,  rigol_thermalization_2008, RevModPhys.83.863, D'Alessio03052016}. As a result, non-thermal, oscillatory dynamics takes place in the many-body scar subspace~\cite{turner_weak_2018,turner_quantum_2018,lin_exact_2019,serbyn_quantum_2021,yao_quantum_2022,moudgalya_quantum_2022,chandran_quantum_2023}. 
On the other hand, the disconnected subspace can thermalize within its small, specific fragment. The resulting Krylov restricted thermalization~\cite{de_tomasi_dynamics_2019,sala_ergodicity_2020,khemani_localization_2020,yang_hilbert-space_2020,Moudgalya_2021,langlett_hilbert_2021,pozsgay_integrable_2021,doggen_stark_2021,rakovszky_statistical_2020,hahn_information_2021,moudgalya_hilbert_2022,moudgalya_quantum_2022,kohlert_exploring_2023,nicolau_local_2023,francica_hilbert_2023,brighi_hilbert_2023,valencia-tortora_rydberg_2024,moudgalya_quantum_2022,aditya_subspace-restricted_2024,ganguli_aspects_2025,yang_probing_2025} has been demonstrated with cold atoms~\cite{adlerObservationHilbertSpace2024,hondaObservationSlowRelaxation2025}, superconducting circuits~\cite{wangExploringHilbertSpaceFragmentation2025} and Rydberg atom arrays~\cite{zhao_observation_2025}. 
In open quantum systems, the interplay between dynamical constraints and dissipation leads to competition between retaining and losing information of initial states, significantly altering the dynamics. Inspired by this perspective, recent studies have shown the stabilization of entangled stationary states in the Temperley-Lieb model~\cite{liHilbertSpaceFragmentation2023}, novel stationary states in the East-West model~\cite{anonymous_exceptional_2025}, and the generation of dissipative time crystals~\cite{liSymmetryinducedFragmentationDissipative2025}. Despite the  growing interest in HSF in the presence of dissipation, analytical insights into experimentally accessible models and the probing of the restricted dynamics with current quantum simulators remain largely unexplored.

\begin{figure}[htbp]
    \centering
    \includegraphics[width=0.92\linewidth]{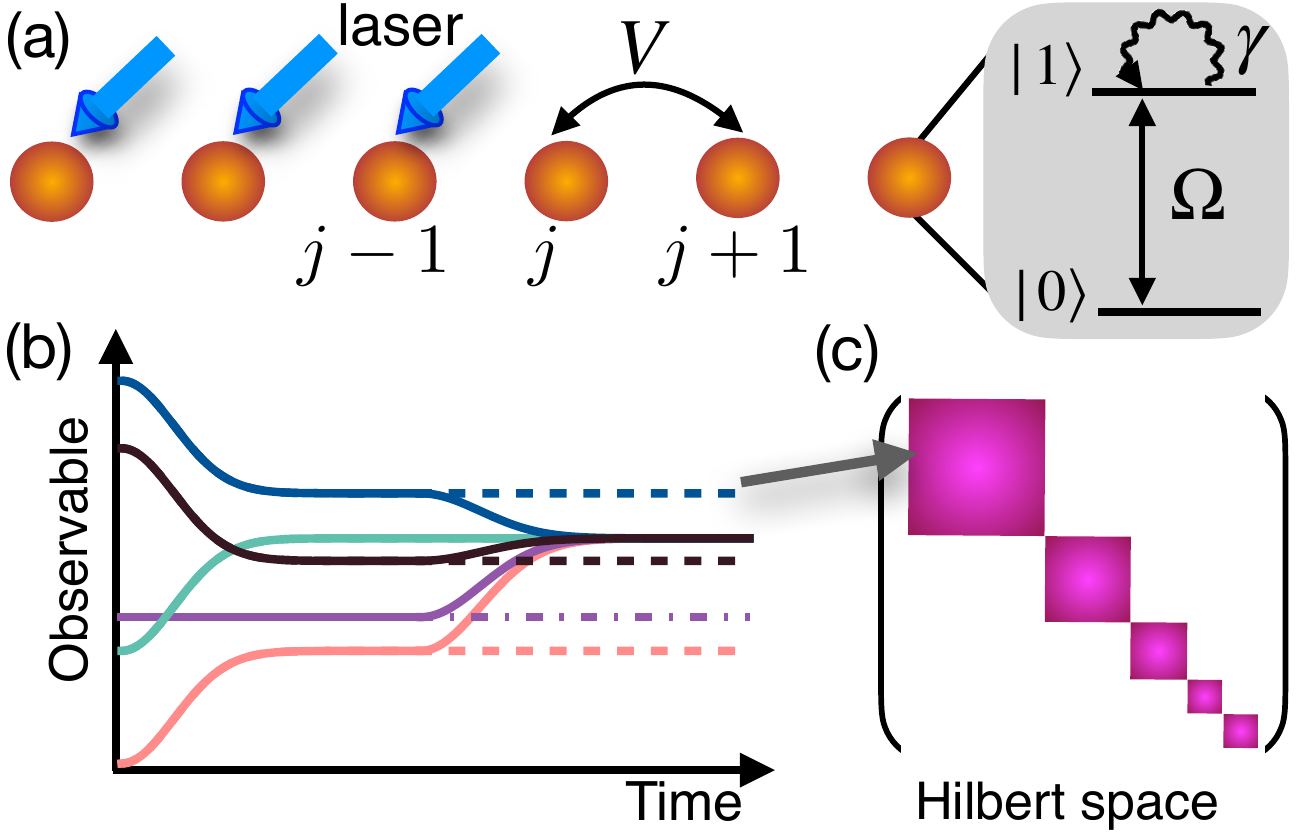}
    \caption{(a) Atom array setting. Atoms are resonantly  excited from the ground state $|0\rangle$ to electronically excited Rydberg state $|1\rangle$ by a laser field (Rabi frequency $\Omega$). In Rydberg state, atoms experience a  dephasing with rate $\gamma$. The van der Waals interaction between Rydberg atoms is dominated by the NN interaction $V$ between the neighboring atoms.  (b) Dynamics of the atom strongly depends on initial states. Typical observable exhibits metastable evolution (solid) in the intermediate timescale before thermalization. The metastable dynamics is effectively described by the dephasing PXP model (dashed). (c) The degenerate zero modes of the dephasing PXP model form distinctive diagonal blocks in the Hilbert space. Dynamical evolution in one block (arrow) is not affected by other blocks.  A frozen state (dot-dashed) is an isolated basis and invariant with time.  See text for details.}
    \label{fig1}
\end{figure}

In this work, we investigate the emergence of HSF in the dynamics 
of a chain of strongly interacting Rydberg atoms under local dephasing, as depicted in Fig.~\ref{fig1}(a). In this driven-dissipative system, different initial states evolve into distinguishable metastable states before thermalizing into the unique stationary state in the long time limit [Fig.~\ref{fig1}(b)]. We show that the multiple metastable states arise from HSF, with their dynamics captured by an effective dephasing PXP model. Uniquely, the effective model supports a large number of degenerate zero modes (stationary states) that form the disconnected Hilbert subspaces [Fig.~\ref{fig1}(c)]. Using a consecutive double excitation addressing operator, we classify the subspaces into classes with conserved symmetries. Dynamics starting from one subspace is entirely confined to the subspace, corresponding to the metastable states of the Rydberg atom chain. We find that numbers of the Hilbert subspaces are determined by a modified Fibonacci sequence, $f(L) = f(L-1) + f(L-2) + f(L-4)$, and grow exponentially with $L$. Our study offers insights into understanding and controlling the HSF in the dissipative regime within  the Rydberg atom array platform. This provides a pathway to explore constrained dynamics, such as strong fragmentation, prethermalization and metastability, across different subspaces.

\textit{Driven-dephasing Rydberg atom chain}--
Our system is a chain of $L$ atoms where the atom is laser addressed individually~\cite{browaeysManybodyPhysicsIndividually2020,ebadiQuantumPhasesMatter2021}. As shown in  Fig.~\ref{fig1}(a), atoms in electronic ground state $|0\rangle$ resonantly couple to Rydberg state $|1\rangle$ with Rabi frequency $\Omega$. In the Rydberg state, the $j$-th and $k$-th atom interact strongly via the van der Waals (vdW) interaction $V_{jk}=V/|j-k|^6$~\cite{saffmanQuantumInformationRydberg2010,shaoRydbergSuperatomsArtificial2024a}, where $V=C_6/a^6$ is the nearest-neighbor (NN) interaction with $C_6$ and $a$ to be the dispersion coefficient and spacing between neighboring sites, respectively. Then dynamics of the atom is described by a many-body spin Hamiltonian  ($\hbar\equiv 1$), ${H}_\text{o} =  \Omega\sum_{j=1}^L   {\sigma}_j^x  + \sum_{j<k}^LV_{jk} {Q}_j {Q}_{k}$, where  $ {\sigma}^x_j = |1\rangle_j\langle 0| + |0\rangle_j\langle 1|$  and  $ {Q}_j = |1\rangle_j\langle 1|$ are the Pauli and projection operators in the $j$-th site. 

In the Rydberg state $|1\rangle$, atoms are subject to local dephasing due to, for instance, laser noise and residual Doppler effect~\cite{honerArtificialAtomsCan2011,helmrichUncoveringNonequilibriumPhase2018}. Density matrix $\rho_{\text{o}}$ of the many-atom system is governed by a Lindblad master equation (ME), $\partial_t {\rho_{\text{o}}} =  {\tilde{\mathcal{L}}}( {\rho_{\text{o}}})$, where Liouvillian ${\tilde{\mathcal{L}}}( {\rho_{\text{o}}})$ reads,
\begin{equation}
     {\tilde{\mathcal{L}}}( {\rho_{\text{o}}}) = -i [ {H}_\text{o},  {\rho_{\text{o}}}] + \mathcal{D}(\rho_{\text{o}}),
    \label{eq:Liouvillian}
\end{equation}
with $\mathcal{D}(\rho_{\text{o}})=\gamma\sum_{j=1}^L\left( {Q}_j  {\rho_{\text{o}}}  {Q}_j^\dagger - \frac{1}{2} \{  {Q}_j^\dagger  {Q}_j,  {\rho_{\text{o}}} \}\right)$  describing the dephasing (rate $\gamma$). The local dephasing eliminates coherence of the system, such that the system equilibrates to the infinite temperature state  $\rho_{\infty}=\mathbb{I}/D$ in the limit $t\to \infty$, where $\mathbb{I}$ and $D=2^L$ are the identity matrix and dimension of the Hilbert space~\cite{siebererUniversalityDrivenOpen2025}. However, dynamics towards the stationarity relies crucially on interplay among the vdW interactions, dephasing, and laser coupling. In the strongly dephasing regime $\gamma \gg |\Omega|$, previous studies have shown critical and universal behaviors in random gases of Rydberg atoms~\cite{lesanovskyKineticConstraintsHierarchical2013a,marcuzziUniversalNonequilibriumProperties2014,hoeningAntiferromagneticLongrangeOrder2014,marcuzziNonequilibriumUniversalityDynamics2015,helmrichSignaturesSelforganizedCriticality2020,kazemiDrivenDissipativeRydbergBlockade2023,siebererUniversalityDrivenOpen2025}. Dynamical constraints driven by the interplay progressively slow down the relaxation processes to the stationary state~\cite{lesanovskyKineticConstraintsHierarchical2013a}. In contrast, we will explore the HSF emerging in the many-atom system in the \textit{strong interaction} regime.

\begin{figure*}[thbp]
    \centering
    \includegraphics[width=0.98\linewidth]{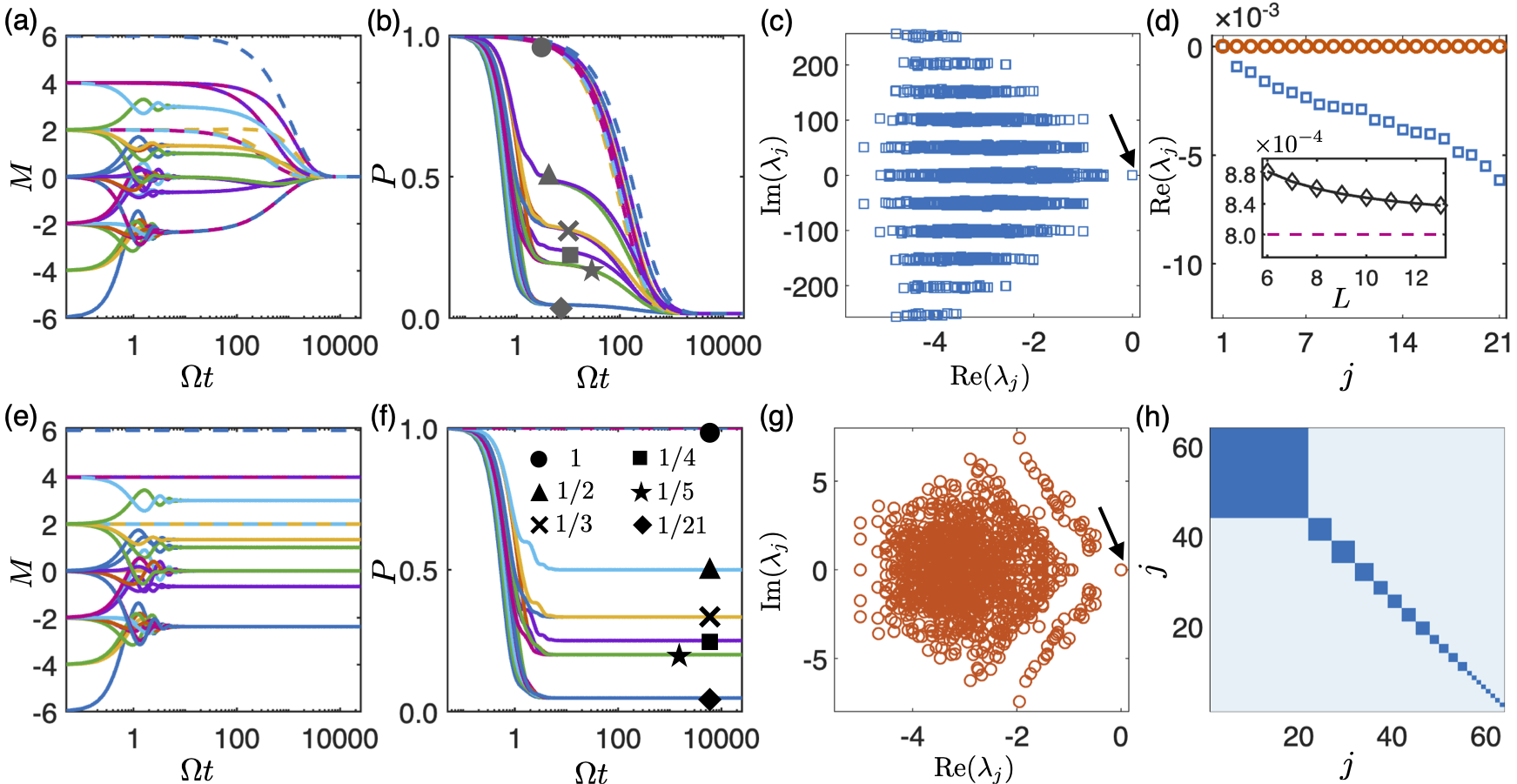}
    \caption{\textbf{Metastable dynamics and fragmentation}. (a) Magnetization $M$ of the Rydberg atom chain. Metastable dynamics takes place from $\Omega t=0$ to $\Omega t\lesssim 100$ (dashed curves) if the initial state is a frozen state. When started with other basis states (solid), the metastability occurs in interval $1\lesssim \Omega t\lesssim 100$. The dependence on the initial state and clustering of $M$ indicate the onset of fragmentation. The dynamics thermalizes to the infinite temperature state $\rho_{\infty}$ in the long time limit. (b) Purity of the atom chain. $\rho_{\text{o}}$ is largely pure when starting from the frozen state (dashed). Other basis (solid) partially retain coherence in the metastable state. (c) Complex spectrum of Liouvillian ${\tilde{\mathcal{L}}}$.  There are 20 nearly degenerate modes (arrow) around the stationary mode ($\lambda_j= 0$). (d) Real parts $\text{Re}(\lambda_j)$ of the quasi-degenerate (square) and degenerate (circle) modes. The inset shows scaling of the gap (minimum of $\text{Re}(\lambda_j)$) with $L$. Increasing $L$, the gap decreases and approaches to the mean field result $\gamma_{\text{mf}}$ (dashed).  (e) Magnetization of the dPXP model. Due to the fragmentation, $M$ does not change with time after the initial transient period (solid). In the frozen state, $M$ is a constant (dashed). (f) Purity of the dPXP model. The steady $P$ is determined by dimensions of the respective subspace, as marked in the figure. In (b), the purity in the metastable state is close to the one of the dPXP model. (g) Complex spectra of the Liouvillian $\mathcal{L}(\rho)$. There are 21 degenerate zero modes, whose real parts are shown in panel (d). (h) Fragmented Hilbert space of the dPXP model. The index $j$ is the decimal number converted from the binary representation of the  basis. Basis states of each subspace are given in Table~\ref{table:HSF}.
    We have considered $V=50\Omega$ in panels (a)-(d), and $L=6$, $\gamma=2\Omega$ in all the panels.} 
    \label{fig2}
\end{figure*}

\textit{Metastable non-thermal dynamics and fragmentation}--
When $V\gg \{\Omega,\ \gamma\}$, a generic behavior of the driven-dephasing dynamics is that memories of initial states are retained for extended periods of time. We illustrate this by investigating dynamics of total magnetization $M=\text{Tr} (\rho_{\text{o}}\sum_j\sigma_j^z)$. Initially, the atom chain is prepared in spin basis states along the $z$-direction, $|s_1 s_2 \cdots s_L\rangle$ with $s_j = 0$ or $1$ denoting spin states in site $j$. At $t=0$, $M=-L,-L+2,\cdots L$, which gives number differences of the spin states in the basis. By using each spin basis as the initial state individually, we numerically solve the ME  and evaluate $M$ for moderate $L$. In Fig.~\ref{fig2}(a), the magnetization of $L=6$ atoms is presented. Two standout features are visible in the dynamics.

First, metastable phases are found in the intermediate timescale~\cite{macieszczakTheoryMetastabilityOpen2016}. In one scenario (dashed curves), $M$ remains constant up to $\Omega t\approx 100$. In the second case (solid), it attains a quasi-stationary value over an interval $1<\Omega t\approx 100$ after a brief transient period ($0<\Omega t \lesssim 1$). In both cases, the magnetization develops plateaus, much longer than the characteristic dephasing time $1/\gamma$, indicating the formation of prethermal phases~\cite{wuPrethermalMemoryLoss2020b,sahaPrethermalizationOpenQuantum2024,artiniCoherentHeatExchange2025}. See more discussions and examples on the metastable dynamics in \textbf{Supplementary materials (SM)}~\cite{SM}. Our analysis shows that in the first scenario (dashed), the dynamics is largely confined to the initial state. As a result, purity of the state, $P=\text{Tr}\rho_{\text{o}}^2$, is high ($P\approx 1$), as shown in Fig.~\ref{fig2}(b). In this case, the initial state is largely a frozen state of the Liouvillian~\cite{breuerTheoryOpenQuantum2007}. Purities of the other initial states (solid) are quasi-stationary when $1\lesssim 1\Omega t\lesssim 100$. Compared to Fig.~\ref{fig2}(a), the metastable period of the purity overlaps with that of $M$.

The metastable timescale is encoded in the complex eigenspectrum $\lambda_j$ of the Liouvillian $\mathcal{L}(\rho_{\text{o}})$. As shown in Fig.~\ref{fig2}(c), most of the eigenmodes have large real parts [$\text{Re}(\lambda_j)> \gamma$], which cause rapid evolution in the early transient period $0<\Omega t \lesssim 1$ [solid curves in Fig.~\ref{fig2}(a) and (b)]. Importantly, we find 20 quasi-degenerate modes around the stationary state [see Fig.~\ref{fig2}(c)], whose $\text{Re}(\lambda_j)$ are much smaller than $\gamma$ [see Fig.~\ref{fig2}(d)]. It is the quasi-degenerate mode that generates the metastable dynamics in Fig.~\ref{fig2}(a) and (b). The lifetime of the metastable dynamics is approximately determined by the  gap (the smallest real part) of the spectra. To obtain the gap, we first carried out mean field analysis, and find an effective rate  $\gamma_{\text{mf}}=\Omega^2\gamma/V^2$~\cite{lesanovskyKineticConstraintsHierarchical2013a,SM}. Apparently, $\gamma_{\text{mf}}\ll \gamma$ in the strong interaction regime $V\gg\Omega$. We also diagonalize the Liouvillian up to $L=13$, and display the gap in the inset of Fig.~\ref{fig2}(d). It is found that the numerical gap is very close to $\gamma_{\text{mf}}$. Another important conclusion drawn from the numerical calculation is the trend that the gap gradually approaches $\gamma_{\text{mf}}$ when $L$ increases. Such scaling indicates that the metastable dynamics is robust even for large system sizes.

Secondly, memory effects play a crucial role in the metastable dynamics. As shown in Fig.~\ref{fig2}(a), trajectories of $M$ combine and form multiple plateaus that are isolated from their neighbors, though some of them can have the same initial values. In this example, there are 9 plateaus in the magnetization and 6 in the purity. These numbers are significantly smaller than Hilbert space dimension $D=64$. 
In the following, we will reveal that the memory effect and clustering of the observable is rooted from the Hilbert (Krylov) space fragmentation~\cite{sala_ergodicity_2020}. The underlying dynamics is captured effectively by a dephasing PXP (dPXP) model.

\begin{table}
    \centering
    \footnotesize
    \renewcommand{\arraystretch}{1.2} 
    \setlength{\tabcolsep}{4.0pt}     
    \begin{tabular}{l r l l p{5.cm}} 
        \toprule
        \textbf{$P$} & M & $\langle \mathcal{A}\rangle$ & \textbf{Rep} & \textbf{Equivalence Class} \\
        \midrule
        $\frac{1}{21}$ & -$\frac{50}{21}$ & 0 & $|000000\rangle$ & $\{|000000\rangle, |000001\rangle, |000010\rangle, |000100\rangle, \newline |000101\rangle, |001000\rangle, |001001\rangle, |001010\rangle, \newline |010000\rangle, |010001\rangle, |010010\rangle, |010100\rangle, \newline |010101\rangle, |100000\rangle, |100001\rangle, |100010\rangle, \newline |100100\rangle, |100101\rangle, |101000\rangle, |101001\rangle, \newline |101010\rangle\}$ \\
        \midrule
        \multirow{2}{*}{$\frac{1}{5}$} 
        & 0 & 2 & $|110000\rangle$ & $\{|110000\rangle, |110001\rangle, |110010\rangle, |110100\rangle, \newline |110101\rangle\}$ \\
        & 0 & 32 & $|000011\rangle$ & $\{|000011\rangle, |001011\rangle, |010011\rangle, |100011\rangle, \newline |101011\rangle\}$ \\
        \midrule
        $\frac{1}{4}$ & 0 & $8$ & $|001100\rangle$ & $\{|001100\rangle, |001101\rangle, |101100\rangle, |101101\rangle\}$ \\
        \midrule
        \multirow{4}{*}{$\frac{1}{3}$}
        & $-\frac{2}{3}$ & 4 & $|011000\rangle$ & $\{|011000\rangle, |011001\rangle, |011010\rangle\}$ \\
        & $\frac{4}{3}$ &6 & $|111000\rangle$ & $\{|111000\rangle, |111001\rangle, |111010\rangle\}$ \\
        & $-\frac{2}{3}$ & 16 & $|000110\rangle$ & $\{|000110\rangle, |010110\rangle, |100110\rangle\}$ \\
        & $\frac{4}{3}$ & 48 & $|000111\rangle$ & $\{|000111\rangle, |010111\rangle, |100111\rangle\}$ \\
        \midrule
        \multirow{4}{*}{$\frac{1}{2}$} 
        & 1 & 12 & $|011100\rangle$ & $\{|011100\rangle, |011101\rangle\}$ \\
        & 3 & 14 & $|111100\rangle$ & $\{|111100\rangle, |111101\rangle\}$ \\
        & 1 & 24 & $|001110\rangle$ & $\{|001110\rangle, |101110\rangle\}$ \\
        & 3 & 56 & $|001111\rangle$ & $\{|001111\rangle, |101111\rangle\}$ \\
        \midrule
        \multirow{9}{*}{1} & 2 & 18 & $|110110\rangle$ & $\{|110110\rangle\}$ \\
        & 2& 28 & $|011110\rangle$ & $\{|011110\rangle\}$ \\
        & 4 & 30 & $|111110\rangle$ & $\{|111110\rangle\}$ \\
        & 2 & 34 & $|110011\rangle$ & $\{|110011\rangle\}$ \\
        & 2 & 36 & $|011011\rangle$ & $\{|011011\rangle\}$ \\
        & 4 & 38 & $|111011\rangle$ & $\{|111011\rangle\}$ \\
        & 4 & 50 & $|110111\rangle$ & $\{|110111\rangle\}$ \\
        & 4 & 60 & $|011111\rangle$ & $\{|011111\rangle\}$ \\
        & 6 & 62 & $|111111\rangle$ & $\{|111111\rangle\}$ \\
        \bottomrule
    \end{tabular}
    \caption{Representative basis and their equivalence classes. Here $P$, $M$, and $\mathcal{A}$ are expectation values of the purity, magnetization, and CDEA operator, correspondingly. The chain length is $L=6$.}
    \label{table:HSF}
\end{table}

\textit{Liouvillian zero modes and Hilbert space fragmentation of the dephasing PXP model}--
To derive the effective model, we make use of a unitary transformation $U=e^{-iVt\sum_jQ_jQ_{j+1}}$. In the strong interaction limit, the transformation leads to the PXP Hamiltonian ${H}_\text{PXP} = \Omega\sum_{j=2}^{L-1}  {P}_{j-1} {\sigma}^x_j {P}_{j+1}$~\cite{PhysRevA.86.041601,PhysRevLett.106.025301,serbyn_quantum_2021}. In deriving ${H}_\text{PXP}$, we have neglected fast oscillating terms and kept only the NN interaction~\cite{SM}. It can be shown that longer range interactions do not affect the underlying HSF, see discussions in \textbf{SM}~\cite{SM}. Using the fact that this transformation does not affect the dephasing operators in Eq.~(\ref{eq:Liouvillian}), we obtain the dPXP model, governed by Liouvillian $ {\mathcal{L}}( \rho) = -i [ {H}_\text{PXP},  {\rho}] + \mathcal{D}(\rho)$ with $\rho = U^{\dagger}\rho_{\text{o}}U$. We will show that the dPXP model not only captures the metastable dynamics, but also provides important insights into the fragmentation.

To illustrate the accuracy, we first calculate the magnetization and purity with the dPXP model, and show the result  in Fig.~\ref{fig2}(e)-(f). Both the magnetization and purity are restricted to special values, which agree with the data shown in Fig.~\ref{fig2}(a) and (b) up to $\Omega t\approx 100$. When $\Omega t>100$, the ones obtained from the dPXP model remain invariant with time, whose values depend only on the initial state. The complex spectra of the dPXP model are then calculated numerically and shown in Fig.~\ref{fig2}(g). Central to the spectra is the presence of 21 degenerate zero modes, whose eigenvalues have zero real parts [Fig.~\ref{fig2}(d)]. Each mode is a maximally mixed state $\rho_d=\mathbb{I}_{d}/d$ in the spin basis representation, where $d$ is the dimension of the respective Hilbert subspace. These zero modes form a series of \textit{disconnected Hilbert subspaces}, giving rising to the HSF. The memory effect arises in the dynamics such that  initial states that fall in different subspaces will not influence each other [see illustration in Fig.~\ref{fig2}(e) and (f)].

The classical nature of the zero modes permits us to construct the Hilbert subspace iteratively. This is done by starting from a basis, and applying $H_{\text{PXP}}$ to the basis repeatedly.  This process identifies basis states linked to the initial state, which collectively form a subspace, also referred to as an equivalence class. For each class, we select a representative basis state that has the smallest decimal value when converted from its binary representation. The procedure is then repeated by selecting a new basis that is not part of a previously identified subspace. This approach explicitly and systematically finds all the equivalence classes. In Table.~\ref{table:HSF}, we present the equivalence classes and their corresponding representatives for $L=6$. With the equivalence classes, we can straightforwardly construct the fragmented Hilbert space of the dPXP model, as shown in Fig.~\ref{fig2}(h).

Both the purity and magnetization can be evaluated once the class is determined. If a class consists of multiple basis states, its stationary purity is determined by the dimension $d$ of the subspace, $P(t\to \infty) = 1/d$, which are listed in Table~\ref{table:HSF} for different classes. There is a special group, where each class contains a single basis [bottom of Table~\ref{table:HSF}]. This basis alone is a zero mode. In other words, they are so-called frozen states in the dynamical evolution, where observable is invariant with time~\cite{breuerTheoryOpenQuantum2007}. For $L = 6$, the 9 frozen states are $|011011\rangle$, $|011110\rangle$, $|011111\rangle$, $|110011\rangle$, $|110110\rangle$, $|110111\rangle$, $|111011\rangle$, $|111110\rangle$, and $|111111\rangle$. Similarly, the magnetization can be determined analytically in each class,  as given in Table~\ref{table:HSF}. The analytical results match that of the numerical result  [Fig.~\ref{fig2}(a) and (b)]. 

\begin{figure}
    \centering
    \includegraphics[width=1\linewidth]{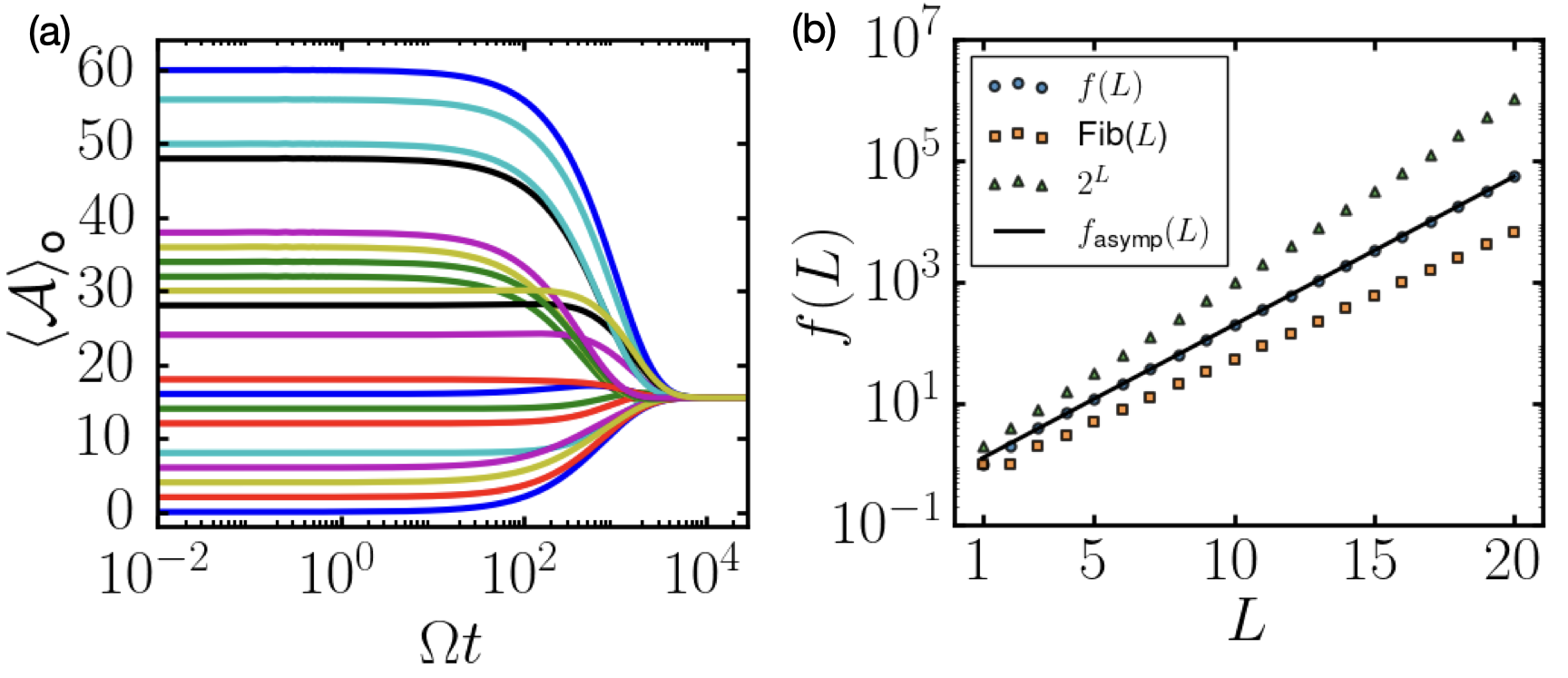}
    \caption{(a) Dynamics of $\langle\mathcal{A}\rangle_{\text{o}}$. In the metastable regime ($\Omega t\le 100$), the CDEA operator is almost constant. Parameters are same with ones in Fig.~\ref{fig2}. (b) Number $f(L)$ of the zero modes with respect to chain length $L$. As shown by the solid line, $f(L)$ increases with $L$ exponentially. For a given $L$, $f(L)$ is larger than the Fibonacci sequence (square), but smaller than the Hilbert space dimension $D=2^L$ (triangle). }
    \label{fig3}
\end{figure}
\textit{Strong symmetry and scaling of the HSF}--
The symmetry of the Liouvillian can be defined with the consecutive double excitations addressing operator (CDEA), 
\begin{equation}
    \label{eq:cdea}
     {\mathcal{A}} = \sum_{j=1}^{L-1} 2^j  {Q}_j  {Q}_{j+1},
\end{equation}
which is the weighted sum of consecutive double excitations (given by adjoint projection operator $Q_jQ_{j+1}$) along the spin chain. Every pair is assigned a unique weight $2^j$ according to its position in the chain (from left to right). The CDEA operator commutes with the PXP Hamiltonian and every dephasing operator, i.e. $[ {H}_\text{PXP},  {\mathcal{A}}]=0$ and $[ {\mathcal{A}},  {Q}_j]=0~\forall j$. Hence it acts as a strong symmetry of the Liouvillian $\mathcal{L}(\rho)$~\cite{liHilbertSpaceFragmentation2023}. In Table~\ref{table:HSF}, the expectation value $\langle 
\mathcal{A}\rangle$ of the equivalence class is given. Each class has a unique expectation value, such that the CDEA operator identifies the symmetry of the fragmented subspace. We point out that the CDEA operator preserves the strong symmetry in the Rydberg atom chain system approximately. In Fig.~\ref{fig3}(a), dynamical evolution of the expectation value $\langle \mathcal{A}\rangle_\text{o}=\text{Tr}(\rho_{o}\mathcal{A})$ is shown. We find that  $\langle \mathcal{A}\rangle_{\text{o}}$ is a constant and coincides with the corresponding $\langle \mathcal{A}\rangle$, before converging to the thermal value $\langle \mathcal{A}\rangle_{\infty}=\text{Tr}(\rho_{\infty}\mathcal{A})=31/2$.

With the help of  Hamiltonian $H_{\text{PXP}}$ and using the commutant algebra~\cite{moudgalya_hilbert_2022}, we are able to obtain the total number $f(L)$ of the subspaces, which follows a recursive relation~\cite{SM},
\begin{equation}
    \label{eq:recursive}
    f(L) = f(L-1) + f(L-2) + f(L-4),
\end{equation}
with initial conditions $f(0) = 1$, $f(-1) = 1$, $f(-2) = 0$, and $f(-3) = -1$. Importantly,  $f(L)$ scales with $L$ exponentially when $L$ is large~\cite{SM},
\begin{equation}
    f_{\text{asymp}}(L) \approx A \alpha^L,
    \label{eq:scaling}
\end{equation}
where $A\approx0.722124$ and $\alpha\approx1.754877$.  As shown in Fig.~\ref{fig3}(b), the scaling agrees with the exact $f(L)$ well. It is different from  the Fibonacci sequence due to the $f(L-4)$ term. In fact, it surpasses scaling of the Fibonacci sequence $\text{F}(L)\approx\varphi^L/\sqrt{5}$ with $\varphi=(1+\sqrt{5})/2$ to be the golden ratio. Despite the exponential scaling, the total number of the subspaces is significantly smaller than the total Hilbert space dimension $D$ [Fig.~\ref{fig3}(b)].

\textit{Conclusion}-- We have investigated the HSF in a chain of strongly interacting Rydberg atoms under local dephasing. The onset of the metastable HSF is captured effectively by the dPXP model. The multiple, degenerate zero modes of the dPXP model form the fragmented Hilbert subspace. Using the CDEA operator, we have classified the zero modes, and identified the basis states that compose the zero modes. The number of the zero modes grows exponentially with the chain length, given by the modified Fibonacci sequence. The exponential growth indicates the increasing complexity of the HSF as the system size increases. This is different from the PXP model, in which the many-body scar states are approximately separated from the remaining Hilbert space. Benefited from the precise control of the coherent~\cite{browaeysManybodyPhysicsIndividually2020,ebadiQuantumPhasesMatter2021,zhao_observation_2025} and dissipative couplings~\cite{chenCollectiveDissipationEngineering2025}, our study paves a route to investigate the dissipative fragmentation and emerging dynamically constrained dynamics, such as entanglement and correlation in the fragmented Hilbert space~\cite{liHilbertSpaceFragmentation2023,moharramipourSymmetryEnforcedEntanglementMaximally2024a,liSymmetryinducedFragmentationDissipative2025}, and competition between localization and fragmentation in the presence of noises (e.g. in Rabi frequency or laser detuning)~\cite{herviouManybodyLocalizationFragmented2021a}, with the state-of-the-art Rydberg atom simulator~\cite{browaeysManybodyPhysicsIndividually2020,ebadiQuantumPhasesMatter2021}.

\textit{Acknowledgments}--We thank Stephen Powell, Juan Garrahan, Masud Haque, Kay Brandner, Arash Jafarizadeh, Jonas Glatthard, Konstantinos Sfairopoulos and Yuqiang Liu for helpful discussion. We acknowledge support from the EPSRC through Grant No. EP/W015641/1 and No. EP/W524402/1, and the use of the University of Nottingham's Ada HPC service.

\bibliography{references}

	\clearpage
\begin{widetext}

	\begin{center}
	{\Large \bf  Supplementary Material: Hilbert space fragmentation in driven-dephasing Rydberg atom array}
	\end{center}
	\vspace{10mm}
	\setcounter{secnumdepth}{2}
	\setcounter{equation}{0}
	\setcounter{figure}{0}
	\setcounter{table}{0}
	\setcounter{page}{1}
	\makeatletter
	
	\renewcommand{\theequation}{S\arabic{equation}}
	\renewcommand{\thefigure}{S\arabic{figure}}
	\renewcommand{\bibnumfmt}[1]{[S#1]}
	%
	
	
	This supplementary material gives further details on the analysis in the main text.
	
	\section{Metastable dynamics for different $L$}
	When numerically solving dynamics of the Rydberg atom chain, the calculation is limited to finite $L$.  With increasing $L$, the required computation memory increases rapidly, as the dimension of the Hilbert space is $D=2^L$. At the same time, the computation time also increases significantly, because the underlying system is stiff, i.e. $V\gg \{\Omega, \gamma\}$. To capture the dynamics accurately, this demands smaller time steps in the numerical simulations in order to mitigate the errors. In the calculations, we have set the time step $\delta t \ll 1/V$, which leads to good convergence in the simulation.
	
	As shown by the inset of Fig.~2(d) in the main text, the lifetime of the metastable phase increases gradually.   In Fig.~\ref{supp_fig3}, we present dynamics of the magnetization and purity for  $L=7$, $8$, and $9$. In all cases, the metastable plateaus persist, and the decay time $\tau$ increases with chain length $L$, consistent with the analysis based on the spectra gap and other calculations. This is important in order to test the scaling with the Rydberg atom quantum simulator.
	
	\begin{figure}[htbp]
		\centering
		\includegraphics[width=0.85\linewidth]{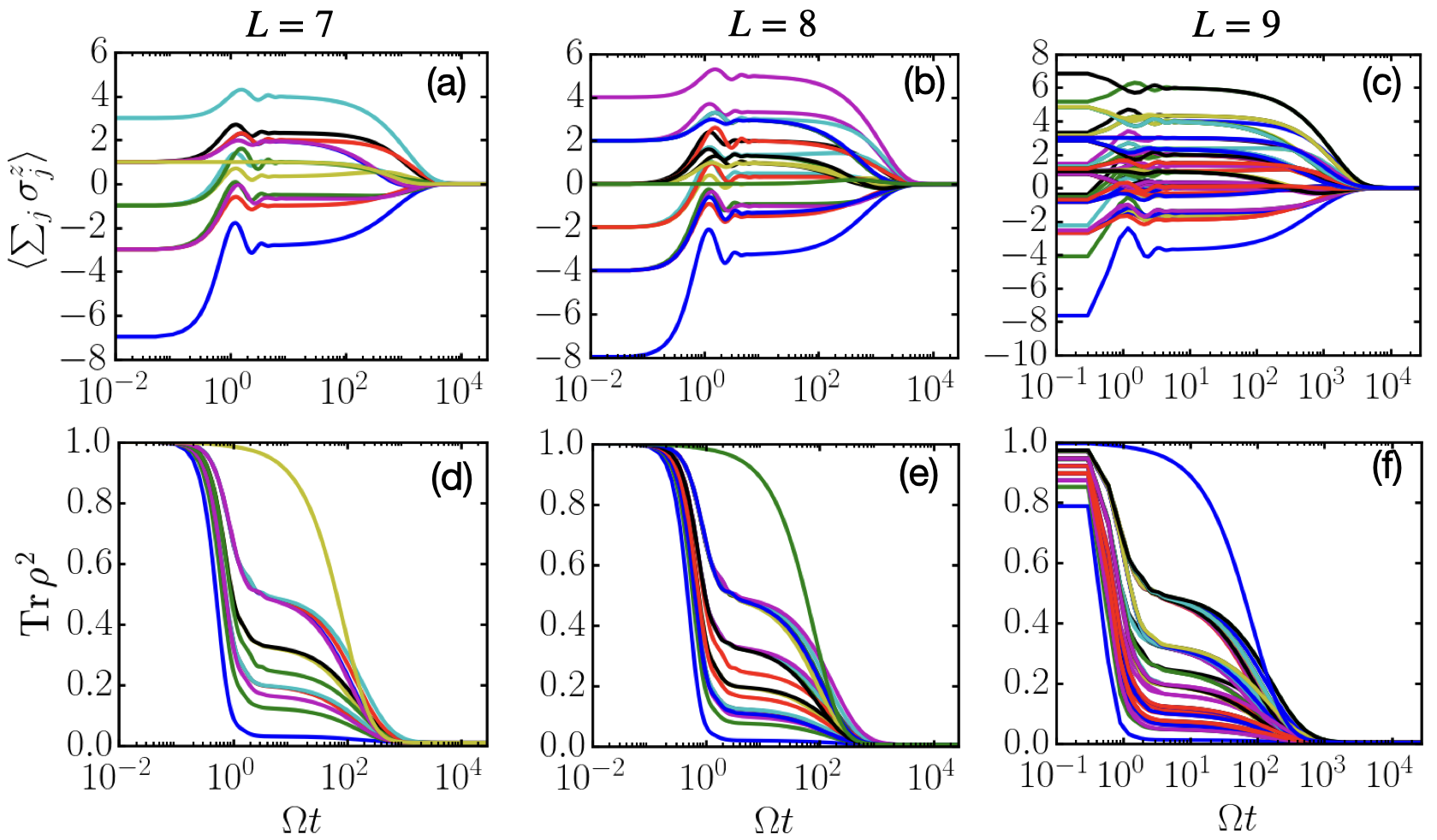}
		\caption{Dynamics of the total magnetization and purity of the Rydberg atom chain with $L=7,\ 8,\ 9$. In all panels, we choose $V=50\Omega$ and $\gamma=2\Omega$.}
		\label{supp_fig3}
	\end{figure}
	\section{Derivation of the effective Liouvillian}
	In the main text, we consider a chain of atoms driven resonantly by a laser from the groundstate to the Rydberg state. In the Rydberg state, atoms experience local dephasing. Dynamics of the atom chain is described by the master equation
	\begin{equation}
		\frac{d {\rho}_{\text{o}}}{dt} = -i [ {H}_\text{o},  {\rho_{\text{o}}}] + \gamma \sum_{j=1}^L\left(  {Q}_j  {\rho_{\text{o}}}  {Q}_j^\dagger - \frac{1}{2} \{  {Q}_j^\dagger  {Q}_j,  {\rho_{\text{o}}} \} \right),
	\end{equation}
	where $ {H}_\text{o} = \Omega \sum_{j=1}^L  {\sigma}_j^x  + \sum_{j<k}^LV_{jk} {Q}_j {Q}_{k}$ is the Hamiltonian of the atoms, and $\rho_{\text{o}}$ is the density matrix of the many-atom system.
	Following the approach used in Refs.~\cite{PhysRevA.86.041601,PhysRevLett.106.025301}, we make a unitary rotation using the nearest-neighbor interaction Hamiltonian $H_1=V\sum_jQ_jQ_{j+1}$,
	\begin{equation}
		{U} = \exp\left(-iVt\sum_{j} {Q}_j {Q}_{j+1}\right). 
	\end{equation}
	In the interaction picture, the density matrix becomes $\rho=U\rho_{\text{o}}U^{\dagger}$. Dynamics of the density matrix $\rho$ is obtained, 
	\begin{equation}
		\frac{d {\rho}}{dt} = -i  [ {U}^{\dagger}({H}_\text{o}-H_1)U,  {\rho}] + \gamma \sum_{j=1}^L  \left(  {U}^{\dagger}{Q}_j{U}  {\rho}  {U}^{\dagger}{Q}_j^\dagger{U} - \frac{1}{2} \{  {U}^{\dagger}{Q}_j^\dagger  {Q}_j{U},  {\rho} \} \right).
	\end{equation}
	As $ {U}$ and $Q_j$ commute, the form of the dephasing is invariant after the transformation, 
	\begin{equation}
		\sum_{j=1}^L  \left(  {U}^{\dagger}{Q}_j{U}  {\rho}  {U}^{\dagger}{Q}_j^\dagger{U} - \frac{1}{2} \{  {U}^{\dagger}{Q}_j^\dagger  {Q}_j{U},  {\rho} \} \right) =  \sum_{j=1}^{L} \left( {Q}_j  {\rho}{Q}_j^{\dagger}-\frac{1}{2}\{ {Q}_j^{\dagger} {Q}_j,  {\rho}\}\right).
	\end{equation}
	Hamiltonian $H'=H_{\text{o}}-H_1 = \Omega \sum_{j=1}^L  {\sigma}_j^x  + \sum_{|j-k|>1}^LV_{jk} {Q}_j {Q}_{k}$ describes the laser-atom coupling and long range interaction (excluding the nearest-neighbor interaction). It is apparent that the two-body interaction part commutes with $U$. Hence its form remains the same after the transformation. The only nontrivial part is the laser-atom coupling term. Applying the transformation, we obtain,
	\begin{equation}
		\begin{split}
			{U}^{\dagger} {H'} {U} &= \Omega\sum_{j=1}^{L} \exp\{iVt( {Q}_{j-1} {Q}_j+ {Q}_{j} {Q}_{j+1})\} {\sigma}^{+}_j \exp\{-iVt( {Q}_{j-1} {Q}_j+ {Q}_{j} {Q}_{j+1})\} + \text{h.c.} + \sum_{|j-k|>1}^LV_{jk} {Q}_j {Q}_k,\\
			&=\Omega\sum_{j=1}^{L} \exp\{iVt( {Q}_{j-1}+ {Q}_{j+1})\}( {P}_{j-1}+ {Q}_{j-1})( {P}_{j+1}+ {Q}_{j+1}) {\sigma}^{+}_j + \text{h.c.}  + \sum_{|j-k|>1}^LV_{jk} {Q}_j {Q}_k,\\
			&=\Omega\sum_{j=1}^{L}( {P}_{j-1}+ {Q}_{j-1}e^{iVt}) {\sigma}^{+}_j( {P}_{j+1}+ {Q}_{j+1}e^{iVt})+\text{h.c.} +  \sum_{|j-k|>1}^LV_{jk} {Q}_j {Q}_k,\\
			&\approx H_{\text{PXP}}+  \sum_{|j-k|>1}^LV_{jk} {Q}_j {Q}_k.
		\end{split}
	\end{equation}
	In the last step, we have neglected the fast rotating term, which is a reasonable approximation in the strong interaction limit $V\gg \{\Omega,\gamma\}$. We point out that the long-range interaction part in the last line commutes with the CDEA operator. Hence it will not affect the  symmetry of the fragmentation. Considering the fact that the vdW interaction decays rapidly with the distance, we therefore ignore this term in the last line. This leads to the dephasing PXP model, 
	\begin{equation}
		\label{se:dephasingPXP}
		\frac{d {\rho}}{dt} = -i [ {H}_\text{PXP},  {\rho}] + \gamma \sum_{j=1}^L\left(  {Q}_j  {\rho}  {Q}_j^\dagger - \frac{1}{2} \{  {Q}_j^\dagger  {Q}_j,  {\rho} \} \right),
	\end{equation}
	with $ {H}_\text{PXP} = \Omega\sum_{j=2}^{L-1}  {P}_{j-1} {\sigma}^x_j {P}_{j+1} +  {\sigma}^x_1 {P}_2 +  {P}_{L-1} {\sigma}_L^x$. With the dephasing PXP model, we have analyzed the spectra gap and the Hilbert space fragmentation.
	
	We remark that we have included the long range interaction terms in the numerical simulations of the Rydberg atom chain, where the nearest-neighbor, next-nearest-neighbor and next-next-nearest-neighbor interactions in $H_{\text{o}}$ are taken into account. We have verified that the other longer range interactions play no roles in the dynamics. It is the strong nearest-neighbor interaction that plays a critical role in determining the metastable dynamics. This is consistent with the effective dephasing PXP model analysis, which is derived under the ideal blockade condition (i.e. including only the nearest-neighbor interaction in the Hamiltonian).

	\section{Mean field approximation and the effective dephasing rate }
	
	To determine the dephasing rate, we consider the mean-field approximation of the master equation by decoupling the many-atom density matrix into the product of the density matrix of individual sites, $\rho \approx \otimes \Pi_j \rho_j$. At site-$j$, we derive dynamics of the diagonal matrix element of local density matrix $\rho_j$,
	\begin{equation}
		\begin{split}
			\frac{d\rho^{(j)}_{rr}}{dt}&=-i\Omega\left(\rho_{gr}^{(j)}-\rho_{rg}^{(j)}\right),\\
			\frac{d\rho_{gg}^{(j)}}{dt}&=i\Omega\left(\rho_{gr}^{(j)}-\rho_{rg}^{(j)}\right),
		\end{split}
		\label{diag_master}
	\end{equation}
	which depends on the off-diagonal matrix elements, 
	\begin{equation}
		\begin{split}
			\frac{d\rho_{gr}^{(j)}}{dt}&=-i\Omega\left(\rho_{rr}^{(j)}-\rho_{gg}^{(j)}\right)+i\Delta_\mathrm{eff}\rho_{gr}^{(j)}-\frac{\gamma}{2}\rho_{gr}^{(j)},\\
			\frac{d\rho_{rg}^{(j)}}{dt}&=-i\Omega\left(\rho_{gg}^{(j)}-\rho_{rr}^{(j)}\right)-i\Delta_\mathrm{eff}\rho_{rg}^{(j)}-\frac{\gamma}{2}\rho_{rg}^{(j)},
		\end{split}
		\label{off_diag_master}
	\end{equation}
	where $\Delta_\mathrm{eff}=V\left(\langle Q_{j-1}\rangle+\langle Q_{j+1}\rangle\right)$. Due to the dephasing, the dynamics of the off-diagonal elements can be adiabatically eliminated. Solving the two coupled equations, we find the stationary solution of the off-diagonal matrix elements, 
	\begin{equation}
		\begin{split}
			\rho_{gr}^{(j)}&=\frac{i\Omega}{i\Delta_\mathrm{eff}-\gamma/2}\left(\rho_{rr}^{(j)}-\rho_{gg}^{(j)}\right),\\
			\rho_{rg}^{(j)}&=\frac{i\Omega}{i\Delta_\mathrm{eff}+\gamma/2}\left(\rho_{rr}^{(j)}-\rho_{gg}^{(j)}\right).
		\end{split}
	\end{equation}
	
	By plugging $\rho_{gr}^{(j)}$ and $\rho_{rg}^{(j)}$ back into Eq.~(\ref{diag_master}),  we obtain the rate equation describing dynamics of the diagonal matrix elements, 
	\begin{equation}
		\begin{split}
			\frac{d\rho_{rr}^{(j)}}{dt} &= -\gamma_{\text{mf}} \left(\rho_{rr}^{(j)}-\rho_{gg}^{(j)}\right),\\
			\frac{d\rho_{gg}^{(j)}}{dt} &= \gamma_{\text{mf}} \left(\rho_{rr}^{(j)}-\rho_{gg}^{(j)}\right).
		\end{split}
	\end{equation}
	Dynamics of the population depends on $\gamma_{\text{mf}}$ given by
	\begin{equation}
		\label{overall_trans}
		\gamma_{\text{mf}} = \Omega^2\frac{\gamma}{\Delta_{\text{eff}}^2+\gamma^2/4},
	\end{equation}
	which is the effective dephasing rate $\Gamma$ that also determines decay of the metastable prethermal phase. In the strong interaction limit and assuming the dynamics  thermalizes ($\langle Q_j\rangle = \frac{1}{2}$), we obtain $\gamma_{\text{mf}}\approx \frac{\Omega^2\gamma}{V^2}$, which depends on the nearest-neighbor interaction, Rabi frequency and single atom dephasing rate. By increasing $|V/\Omega|$, the effective dephasing rate decreases, i.e. the lifetime of the metastable phase is increased.
	
	To verify the validity of the rate $\gamma_{\text{mf}}$, we have numerically solved the dynamics of 6 atoms using Hamiltonian $H_{\text{o}}$. We then find the lifetime of the metastable phase when the initial state is the basis state $|000000\rangle$, i.e. all atoms are in the groundstate. We vary $\gamma$ while fixing $V$ and $\Omega$. From the numerical data, we obtain the lifetime, shown in Fig.~\ref{supp_fig1}(a). The effective dephasing rate $\Gamma$ agrees well with the numerical data (dots).
	
	\section{Scaling of length of prethermalization plateau with respect to $\gamma$}
	\begin{figure}
		\centering
		\includegraphics[width=0.9\linewidth]{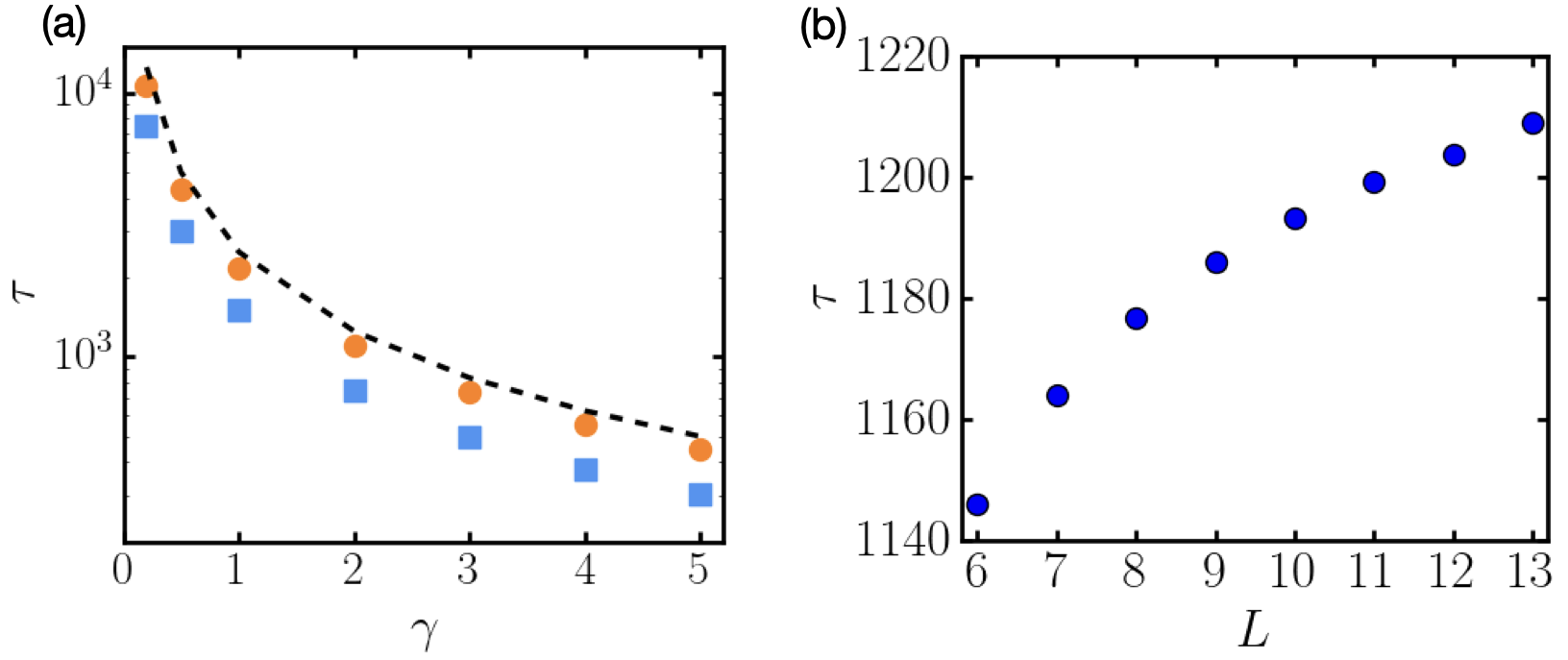}
		\caption{(a) Decay time $\tau$ versus dephasing strength $\gamma$. The black dashed line stands for Eq.~(\ref{overall_trans}). The orange points are taken from the  spectra gap of the Liouvillian $\mathcal{L}(\rho_{\text{o}})$. The square points are obtained from the time when the dynamics of total magnetization reaches $1/e$ of the  steady state computed with the dPXP model. In both cases, the initial state is chosen as $|000000\rangle$. (b) Decay time $\tau$ versus chain length $L$. Data points are  the Liouvillian gap of the dPXP model, i.e. the smallest value along the real axis near the zero mode. Other parameters are $V=50\Omega$ and $\gamma=2\Omega$ in panel (b).}
		\label{supp_fig1}
	\end{figure}

	Here, we study the scaling of the decay time $\tau$ with dephasing strength $\gamma$. In the previous section, we showed that in the strong interaction limit $V \gg \gamma$, the characteristic dephasing time is
	\begin{equation}
		\label{decay_rate}
		\tau = \frac{1}{\gamma_{\text{mf}}} 
	\end{equation}
	In Fig.~\ref{supp_fig1}(a), we plot $\tau$ versus $\gamma$ for $L=6$, $\Omega=1$, $\Delta=0$, $V=50$, and initial state $|000000\rangle$. Data points are from the smallest gap along the real axis in the Liouvillian spectrum of the full model near the zero mode, as depicted in Fig.~(2) in the main text. Equation~(\ref{decay_rate}) accurately captures the relation between the decaying time computed from the Liouvillian gap and dephasing strength $\gamma$.
	
	In Fig.~\ref{supp_fig1}(b), we plot $\tau$ versus the chain length $L$ for $\Omega=1$, $\Delta=0$, $V=50\Omega$, and $\gamma=2\Omega$. The data points are obtained from the smallest Liouvillian gap along the real axis near the zero mode, computed using the full model. We observe that $\tau$ increases with $L$. As $L \to \infty$, it presumably approaches the value predicted by Eq.~(\ref{decay_rate}), which for $\gamma=2$ yields $\tau_\infty = 1250$.
	
	\section{Commutant algebra and the recursive relation of $f(L)$}
	We first define the number operator of consecutive double excitations $ {N}=\sum_{j} {Q}_j {Q}_{j+1}$, whose eigenspace reads, $\mathcal{H}_N=\text{span}\{|\overrightarrow{n}\rangle: {N}|\overrightarrow{n}\rangle=N|\overrightarrow{n}\rangle\}$ for $N=0,1,...,L-1$. The CDEA operator $ {\mathcal{A}}=\sum_{j}2^j {Q}_j {Q}_{j+1}$ commutes with $ {N}$ and therefore preserves the number of consecutive double excitations. Thus, $ {\mathcal{A}}$ is a block-diagonal matrix,
	\begin{equation}
		{\mathcal{A}}=\bigoplus_N  {\mathcal{A}}_N,\  {\mathcal{A}}_N|\overrightarrow{n}\rangle=\left(\sum_{j}2^j\delta_{n_j,1}\delta_{n_{j+1},1}\right)|\overrightarrow{n}\rangle,\ \forall |\overrightarrow{n}\rangle \in \mathcal{H}_N.
	\end{equation}
	In other words, each $ {\mathcal{A}}_N$ is also diagonal in $\mathcal{H}_N$. 
	The dissipative term with local dephasing preserves the block-structure. Consequently, the Hamiltonian is block-diagonal:
	\begin{equation}
		{H}=\bigoplus_N  {H}_N,
	\end{equation}
	with each block satisfying $[ {H}_N,  {\mathcal{A}}_N]=0$, ensuring the strong symmetry conditions: $[ {H}, {\mathcal{A}}]=0,\ [ {L}_j,  {\mathcal{A}}]=0$ for all jump operators $ {L}_j$.
	
	\begin{figure}[h!]
		\centering
		\includegraphics[width=1\linewidth]{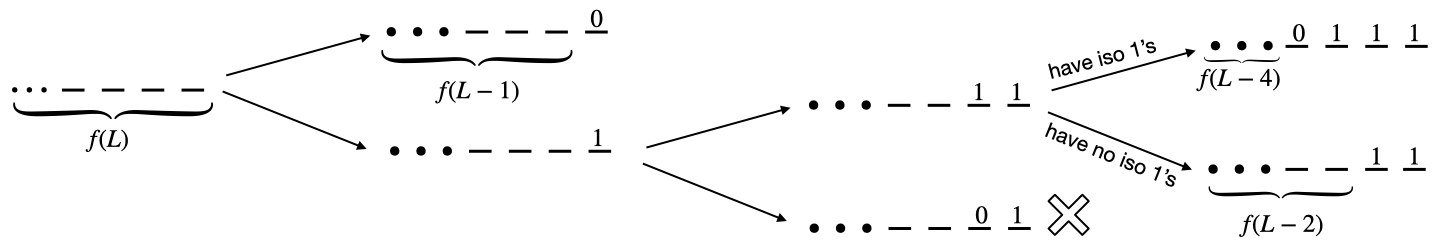}
		\caption{Diagram to identify the recursive relation for $L$ sites. The key step is to count computational basis states with no isolated 1's (i.e., no configurations like $|\dots010\dots\rangle$ in the bulk or $|10\dots\rangle$, $|\dots01\rangle$ at boundaries).}
		\label{supp_fig2}
	\end{figure}
	
	The dimension of each Hilbert subspace $\mathcal{H}_N$ of Hamiltonian ${H}_\text{PXP}$ can be determined using the commutant algebra of the Liouvillian $\mathcal{L}$~\cite{anonymous_exceptional_2025, moudgalyaHilbertSpaceFragmentation2022}. The commutant algebra of $\mathcal{L}$ is spanned by all operators that simultaneously commute with $ {H}_{\text{PXP}}$ and all jump operators $ {Q}_j$. We notice that a general operator $O$ commuting with all dephasing operators $ {Q}_j$ is a diagonal matrix expressed in the computational basis. Cares have to be taken in order to make the operator commuting with $H_{\text{PXP}}$. We now illustrate this using a simple case with $L = 2$. The Hamiltonian becomes $ {H}_{\text{PXP}} =  {X}_1  {P}_2 +  {P}_1  {X}_2$. The operator $ {O}$ is given by a general form 
	\begin{equation}
		{O} = a |11\rangle\langle11| + b |01\rangle\langle01| + c |10\rangle\langle10| + d |00\rangle\langle00|,
	\end{equation}
	where $a, b, c, d \in \mathbb{C}$, and will be determined shortly. If we look at the commutation relation of the individual elements, the first term commute with $H_{\text{PXP}}$. However the other three terms (consisting of basis $|01\rangle$, $|10\rangle$ and $|00\rangle$) are not commuting with $H_{\text{PXP}}$ individually. The constraints $[ {H}_{\text{PXP}},  {O}] = 0$ and $[ {O},  {Q}_j] = 0$ for all $j$ impose that $b = c = d$ in the block of the relevant basis. Thus, for $L = 2$, the commutant algebra $\mathcal{C}$ of $\mathcal{L}$ is obtained,
	\begin{equation}
		\mathcal{C} = \text{Span} \{ |11\rangle\langle11|, (|01\rangle\langle01| + |10\rangle\langle10| + |00\rangle\langle00|) \},
	\end{equation}
	indicating two irreducible blocks and, hence, two steady states for $L = 2$. One irreducible block consists solely of the frozen state (i.e. $|11\rangle$), while the other comprises three basis states, yielding a dimension of 3. The procedure to expand the commutant algebra from 2 to $L>2$ atoms is similar. In the main text,  we have list the block basis that consist of the commutant algebra for $L = 6$ in Table I. We want to point out that the commutant algebra $O$ constructed this way is an Abelian commutant algebra, as each individual terms commute with each other~\cite{moudgalyaHilbertSpaceFragmentation2022}.
	
	\begin{table}[htbp]
		\centering
		\begin{tabular}{|c|*{20}{c|}}
			\hline
			$L$ & 1 & 2 & 3 & 4 & 5 & 6 & 7 & 8 & 9 & 10 & 11 & 12 & 13 & 14 & 15 & 16 & 17 & 18 & 19 & 20 \\
			\hline
			$f(L)$ & 1 & 2 & 4 & 7 & 12 & 21 & 37 & 65 & 114 & 200 & 351 & 616 & 1081 & 1897 & 3329 & 5842 & 10252 & 17991 & 31572 & 55405 \\
			\hline
			$\text{Fib}(L)$ & 1 & 1 & 2 & 3 & 5 & 8 & 13 & 21 & 34 & 55 & 89 & 144 & 233 & 377 & 610 & 987 & 1597 & 2584 & 4181 & 6765 \\
			\hline
		\end{tabular}
		\caption{Sequence comparisons between the variant Fibonacci and Fibonacci sequence up to $L=20$.}
		\label{seq_compare}
	\end{table}
	
	To find the dimension of the commutant algebra i.e., the number of degenerate zeros of $\mathcal{L}$, we notice that each block ${H}_N$ possesses a zero mode, so the number of zero modes of $\mathcal{L}$ depends solely on the number of partitions of $ {H}_{\text{PXP}}$, which corresponds to the number of partitions in $ {\mathcal{A}}$. The number of partitions in $ {\mathcal{A}}$ is determined by counting the distinct configurations with no isolated excitations (i.e., configurations containing $|...010...\rangle$ in the bulk or $|10...\rangle$ and $|...01\rangle$ at the boundaries). In the presence of isolated excitation, this corresponds to the number of substring configurations with no isolated excitations. 
	
	For $L$ sites, as shown in Fig.~\ref{supp_fig2}, we construct these states starting from the rightmost site, which can be 0 or 1. If the rightmost site is 0, the first $L-1$ sites yield $f(L-1)$ states with no isolated 1's. If the rightmost site is 1, the second-rightmost site must be 1 to avoid an isolated 1 at the boundary; in this case, the first $L-2$ sites can either have no isolated 1's, leading to $f(L-2)$ possibilities, or possesses an isolated 1 at their right boundary, in which case the first $L-4$ sites contribute $f(L-4)$ possibilities with no isolated 1's. Thus, the total number of such computational basis states for $L$ sites yields 
	\begin{equation}
		\label{recursive_supp}
		f(L) = f(L-1) + f(L-2) + f(L-4),    
	\end{equation}
	with initial conditions $f(0) = 1$, $f(-1) = 1$, $f(-2) = 0$, and $f(-3) = -1$. $f(L)$ gives the number of degenerate zero modes of $\mathcal{L}$, which is a modified Fibonacci number. 
	
	To compare with the Fibonacci sequence, in Table~\ref{seq_compare}, we list the results of both the Fibonacci and the modified sequences up to $L=20$. The latter grows significantly faster than the original Fibonacci sequence.

	\section{Scaling of the modified Fibonacci sequence}
	\label{derivation_scaling}
	To determine the asymptotic behavior of the recursive relation $f(L)$ of order $k$, given by
	\begin{equation}
		\label{general_recursive}
		f(L) = a_1 f(L-1) + a_2 f(L-2) + \dots + a_k f(L-k),
	\end{equation}
	where $a_j$ are constants, we assume a trial solution $f(L) = r^L$, with $r \in \mathbb{C}$. Substituting into Eq.~(\ref{general_recursive}) and dividing by $r^{L-k}$ yields the characteristic equation:
	\begin{equation}
		r^k - a_1 r^{k-1} - a_2 r^{k-2} - \dots - a_k = 0.
	\end{equation}
	The roots $r_1, r_2, \dots, r_k$ give the solution,
	\begin{equation}
		f(L) = A_1 r_1^L + A_2 r_2^L + \dots + A_k r_k^L,
	\end{equation}
	where $A_1, A_2, \dots, A_k$ are constants determined by the initial conditions. For the specific relation in Eq.~(\ref{recursive_supp}), the characteristic equation reads,
	\begin{equation}
		r^4 - r^3 - r^2 - 1 = 0.
	\end{equation}
	Factoring as $(r+1)(r^3 - 2r^2 + r - 1) = 0$, we obtain four roots,
	\begin{equation}
		\begin{cases}
			\alpha &\approx 1.754877, \\
			\beta &= -1, \\
			\gamma_{\pm} &= 0.122561 \pm 0.744862i,\ |\gamma_{\pm}| = 0.755 < 1.
		\end{cases}
	\end{equation}
	Thus, the general solution is,
	\begin{equation}
		f(L) = A_1 \alpha^L + A_2 \beta^L + A_3 \gamma_{+}^L + A_4 \gamma_{-}^L.
	\end{equation}
	with $A_1 \approx 0.722124$, $A_2=0$, $A_3\approx0.13894+0.20225i$ and $A_4\approx0.13894-0.20225i$ determined by the initial conditions.
	
	For large $L$, the complex $\gamma$ term decays since $|\gamma| < 1$. The $\alpha$ term dominates, yielding the asymptotic behavior,
	\begin{equation}
		f(L) \approx \ A_1 \alpha^L,
	\end{equation}
	which reproduces Eq.~(4) in the main text.
		\end{widetext}
\end{document}